\pdfoutput=1 
\documentclass{JINST}

\usepackage{url}
\usepackage[Symbol]{upgreek}

\title{On the Single-Photon-Counting (SPC) modes of imaging using an XFEL source}

\author{Zhehui Wang\\ 
Los Alamos National Laboratory,\\
  Los Alamos, NM 87545, U.S.A.\\
E-mail: \email{zwang@lanl.gov}}

\abstract{The requirements to achieve high detection efficiency (above 50\%) and gigahertz (GHz) frame rate for the proposed 42-keV X-ray free-electron laser (XFEL) at Los Alamos are summarized. Direct detection scenarios using C (diamond), Si, Ge and GaAs semiconductor sensors are analyzed. Single-photon counting (SPC) mode and weak SPC mode using Si can potentially meet the efficiency and frame rate requirements and be useful to both photoelectric absorption and Compton physics as the photon energy increases. Multilayer three-dimensional (3D) detector architecture, as a possible means to realize SPC modes, is compared with the widely used two-dimensional (2D) hybrid planar electrode structure and 3D deeply entrenched electrode architecture. Demonstration of thin film cameras less than 100-$\mu$m thick with onboard thin ASICs could be an initial step to realize multilayer 3D detectors and SPC modes for XFELs.}

\keywords{MaRIE XFEL; GHz hard X-ray imaging detector; Single-photon-counting (SPC) mode; Multilayer 3D detector architecture}

\begin{document}

\section{Introduction}\label{sec:intro}

X-ray free electron lasers (XFELs) have opened up new vistas to the mesoscopic (micro-meter length scale) world of structural biology, chemistry, material sciences, geology, etc.~\cite{Miao:1999,RH:2009,Zigler:2011,SC:2014}. With the number of photons reaching 10$^{12}$ to 10$^{13}$ per pulse (10$^{10}$ - 10$^{11}$ photons per $\mu$m$^2$), the spatial resolution down to nano-meters for micro-meter-size polycrystals or non-periodic structures such as biological macromolecules is feasible; {\it i.e.}, the ratio of the sample size to spatial resolution can exceed $10^3$. With individual photon pulses lasting a few to 100 fs, atomic motion is essentially frozen during each photon pulse; {\it i.e.}, the motional blur for atoms moving at approximately 1 km/s (room temperature) is essentially eliminated (0.1 nm or less). With the transverse photon coherence reaching 10 - 100 $\mu$m, X-ray photon correlation spectroscopy (XPCS)~\cite{Sutton:1991}, time-resolved Phase Contrast Imaging (PCI)~\cite{Wilkins:1996,Luo:2012}, and other techniques are being developed or used to track `4D' (1D temporal + 3D spatial) evolution of mesoscopic structures.

Detection of X-ray intensity as a function of the angle of scattering (and time for temporal information) is a common experimental approach, although the X-ray energy, the pulse-to-pulse time delay, the beam intensity, the distance from the sample to detection, and other experimental parameters may vary under different conditions. 
For a characteristic scattering sample length $a_0$, the coherent scattering angle ($\theta$) is given by $m \lambda = a_0 \sin\theta$, with the integer $m \ge 1$ being the order of interference and $\lambda$ the X-ray wavelength. Since $a_0 \gg \lambda$ for hard X-rays, small-angle X-ray scattering (SAXS) is widely used and the scattered X-rays are captured near the direction of beam exit. In addition, the coherently scattered X-ray is more intense than incoherent scattering at small angles. Despite the large peak intensities of XFELs, many repetitive pulses may still be needed to collect enough scattered X-rays for imaging, since individual electron scattering cross section is only about 0.665 barn. The minimum number of X-ray photons to form an image has been shown to depend on the resolution ($\delta$) as $\delta^{-3}$ (partially coherent or incoherent sources) or $\delta^{-4}$ (coherent sources)~\cite{HD:1955,Marchesini:2003,Shen:2004,Howells:2009}, and amounts to about 10$^{8}$ scattered photons per image~\cite{Barber:2014}. Pixelated area detectors, together with computation, are indispensable to XFEL imaging~\cite{Gruner:2002,Falus:2004,CSPAD:2012,AGIPD:2014}.  

 
High-performance pixelated area detectors have been identified as one bottle neck for XFEL~\cite{nxd:rpt}, in particular for X-ray photon energies above 20 keV. XFEL sources and applications drive the further development of pixelated area detectors.
 
\subsection{A 42-keV XFEL source}\label{sec:source}
A new 42-keV XFEL source is proposed by Los Alamos to examine granular matter of high-Z materials. Some of its properties are compared with LCLS-I, SASE1 of the European XFEL (EXFEL), and SACLA in Japan~\cite{pri:Dinh,Arthur:2002,Altarelli:2007,PS}. In addition to higher X-ray energy at 42 keV (fundamental mode), Another unique feature (from the detector perspective) is multiple pulses with a pulse-to-pulse time delay around 1 ns or less.

\begin{table}[hbtp]
\caption{A comparison of key parameters for several XFELs~\cite{pri:Dinh}.}
\label{tab:XFEL}
\smallskip
\centering
\begin{tabular}{|lccccc|}
\hline
Parameters & unit& LCLS-I & EXFEL (SASE1) & SACLA & MaRIE\\
\hline\hline
Electron energy & (GeV) &13.6 &17.5& 8.5 &12 \\
Electrons per pulse & (nC) & 0.25 & 1.0 & 0.2 &0.1\\
SASE gain length & (m) & 3.1& 4.0 & 2.8 & 2.6 \\
Photon energy & (keV) & 8.2 & 12.5 & 11.9 &42\\
Photon pulse length & (fs) & 73.5 & 200 & 100 & 33 \\
Photons per pulse& ($\times$10$^{11}$)& 8.7 & 25 & 1.8 & 0.39 \\
Transverse spot size & ($\mu$m)& 33 & 31 & 28 & 13 \\
Beam divergence & ($\mu$rad) & 0.36 & 0.26 & 0.29 & 0.19\\
\hline
\end{tabular}
\end{table}

\subsection{Detector requirements}\label{sec:req}
The spatial ({\bf r}), temporal ($t$), momentum ({\bf p}) and polarization ({\bf s}) distributions of the scattered X-ray photons determine the imaging detector requirement. In other words, an ideal detector measures the X-ray photon number distribution function $N_\nu({\bf p}, {\bf r}, {\bf s},t)$. In XFEL setting, the directions of the scattered photon ($\theta_0$, $\phi_0$) is well defined after the relative location in-between the detector and the object, $z_0$, is chosen, and the function reduces to  $N_\nu(E_\nu, {\bf r}, {\bf s},t; z_0, \theta_0, \phi_0)$ with $E_\nu = h\nu$ being the photon energy. A pixelated digitial area detector discretizes the measurement both in time and in space, and the measured function is  $N_\nu(E_\nu, k \Delta_x, l \Delta_y, m \Delta_t; z_0,\theta_0, \phi_0)$, here we ignore the polarization-dependence in the measurement. The detector requirements define the energy sensitivity, pixel resolutions $\Delta_x$, $\Delta_y$, temporal resolutions $\Delta_t$, The total number of pixels in the $x-$, $y-$ direction and in time are given by $1 \leq k \leq M_x$, $1 \leq l \leq M_y$, $1 \leq m \leq M_t$ (the discrete storage units for each pixel) respectively. By definition, $0 \leq N_\nu \leq N^{max}$. The detector dynamic range, or the ratio of the maximal photon sensitivity across the whole image divided by the minimum photon sensitivity, is $N^{max}/N^{min}$. Experimentally, it has been found that single X-ray photon sensitivity is required, $N^{min}$ = 1; therefore, $N^{max}$ also defines the dynamic range of the detector. For a single pixel, the number of photons expected is given by
\begin{equation}
N_\nu = \Gamma_0 r^2_e \int |F (\theta_0,\phi_0 )|^2 d\Omega,
\label{eq:nu0}
\end{equation}
where $\Gamma_0$ is the number of incident X-ray photons per unit area per pulse at the sample. $r_e = 2.818 \times 10^{-15}$ m is the classical electron radius. the integration is over the individual pixel photon collection solid angle. If multiple pulses are collected for a single image, $\Gamma_0 = \int dt I_0/(h\nu) $ is the time-integrated photon counts, or time-integrated intensity $I_0$ divided by the photon energy $h\nu$ for a mono-energetic or narrow-bandwidth XFEL. The function $F(\theta_0, \phi_0)$ is called X-ray scattering form factor~\cite{Warren:1990,Guinier:1994}, which is a complex number in general and its modulus determines the scattered X-ray distribution as a function of angles $\theta_0$ and $\phi_0$. The largest photon count happens at the peak of the scattering form factor. $|F(\theta_0, \phi_0)|$ always peaks in the forward direction ($\theta_0 = 0$) and for coherent scattering, it is equal to the total number of electrons in a coherent scattering volume. For a volume that contains $N_0$ atoms and each has $Z$ electrons on average, $|F^{max}(\theta_0, \phi_0)| \leq N_0 Z$. 

Smaller the pixels are, more difficult it becomes to fabricate using micro-processing technologies such as CMOS for integrated circuits. The maximum allowed pixel sizes $\Delta_x = \Delta_y$ are determined by the oversampling requirement in lensless diffractive imaging~\cite{Sayre:1980},
\begin{equation}
\Delta_x =\Delta_y = \frac{z_0\lambda}{ O a_0} = \frac{a_0}{ON_F},
\label{pix:sz}
\end{equation}
here $O \sim 2$ is the so-called oversampling parameter~\cite{Miao:1998}. $a_0$ is the sample size (the diameter of a sphere), and $\lambda$ the wavelength of the X-ray photon. The Fresnel number ($N_F$) is the same as the definition in optics, $N_F = a_0^2/z_0\lambda$. From Eq.~(\ref{pix:sz}), the pixel solid angle $\Delta \Omega  = \Delta_x \Delta_y/ (4 \pi z_0^2) $. Therefore, it seems that the maximum number of photon might be $N^{max} \sim \Gamma_0 r^2_e N_0^2 Z^2  \Delta_x \Delta_y/ (4 \pi z_0^2) $. However, this is an overestimate. Although the peak intensity scales with the number of scatterers as $N_0^2$, the  peak width narrows as $N_0$ increases~\cite{Warren:1990,Guinier:1994}. Correspondingly, the angular spread of the peak is 
\begin{equation}
\Delta \theta^{max} = \frac{\lambda}{4a_0 N_0^{2/3}}.
\label{angular:1}
\end{equation}
Comparison of Eqs.~(\ref{pix:sz}) and (\ref{angular:1}) indicates that the solid angle of a maximum pixel size that satisfies oversampling condition is much greater than the angular spread of a peak, since $N_0 \gg 1$. Therefore, $\int |F (\theta_0,\phi_0 )|^2 d\Omega = N_0 Z$, and the expected peak photon count is
\begin{equation}
N^{max} = \Gamma_0 r^2_e N_0 Z.
\end{equation}

 We compare $N^{max}$ under two different scenarios, $N_F \ge 1$ (the Fresnel regime) and $N_F \ll1$ (the far field or Fraunhoffer regime). The main difference is in $N_0$, the number of atoms constructively interfere with each other and reinforce the scattering peak.
\begin{equation}
N_0 = \left\{ \begin{array}{ll}
\frac{\pi}{6} \frac{\rho_0}{M_0} a^3, &  \hspace{0.2 cm} N_F \geq 1 \\
\frac{\pi}{6} \frac{\rho_0}{M_0} a_0^3,  & \hspace{0.2 cm} N_F \ll 1
\end{array}  \right .,
\label{eq:N0}
\end{equation}
where $a^2 = a_0^2/N_F$. $\rho_0$ is the mass density and $M_0$ the average atomic mass per atom. Eq.~(\ref{eq:N0}) basically means that in the Fraunhoffer regime, the whole volume contributes to the main peak, while in the Fresnel regime, only a fraction of the object contribute to the peak constructively and the fraction is determined by $N_F$.

The overall detector sizes, or the total number of pixels $M_x$, $M_y$, are determined by resolution ($\delta$),
\begin{equation}
M_x\Delta_x = M_y \Delta_y = \frac{\lambda z_0}{\delta} = \frac{a_0^2}{N_F \delta}.
\label{pix:num}
\end{equation}
Combining Eqs.~(\ref{pix:sz}) and (\ref{pix:num}) gives,
\begin{equation}
M_x = M_y = \frac{Oa_0}{\delta}.
\end{equation}

\begin{figure}[tbp] 
\centering
\includegraphics[width=.4\textwidth]{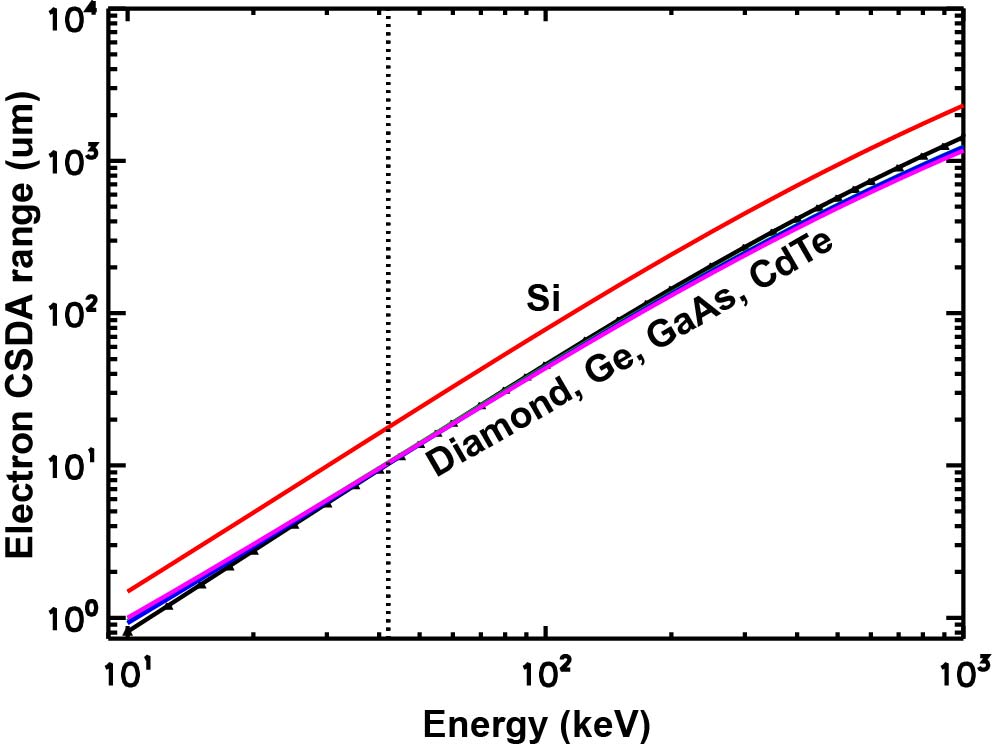}
\caption{CSDA ranges of photoelectron and other secondary electrons as a function of their energies in several semiconductors that are likely used for hard X-ray imaging: C (Diamond), Si, Ge, GaAs and CdTe. The CSDA ranges set a lower limit on the the pixel sizes $\Delta_x$ and $\Delta_y$ at 42 keV.}
\label{fig:CSDA}
\end{figure}

The spatial resolution of the detector is also limited by the stopping or ranges of photoeletrons, Auger electrons and Compton electrons. The continuous slowing-down approximation (CSDA) electron ranges at above 1 keV are available from NIST~\cite{NIST}. The CSDA range with an error up to 10\% is shown in Fig.~\ref{fig:CSDA} for several semiconductors. At 42 keV, the CSDA ranges of electron are 10.2, 17.8, 10.3, 10.5 and 10.4 $\mu$m in C (diamond), Si, Ge, GaAs, and CdTe respectively. We conclude that the pixel size can not be less than the CSDA range of electrons. This constraint is taken into account to define $\Delta_x$ and $\Delta_y$ for 42 keV X-rays in the Table~\ref{tab:detector}. 

\begin{table}[hbtp]
\caption{A comparison of detector designs for a 42 keV ($\lambda = 0.029$ nm) XFEL.}
\label{tab:detector}
\smallskip
\centering
\begin{tabular}{|llccc|}
\hline
Sample size & $a_0$($\mu$m) &0.1 & 2.0 & 20\\
Resolution &$\delta$ (nm) & 2 & 10 &20 \\
Pixel size & $\Delta_x$=$\Delta_y$ ($\mu$m) & 25 & 100 & 100 \\
Fresnel number & $N_F$ & 0.002 & 0.01 & 0.05 \\
Number of pixels & $M_x = M_y$ & 100 & 400 & 1000 \\
Number of frames & $M_t$ & 20 & 100 & 100 \\
Photon flux density & ($\mu$m$^{-2}$) & 10$^{10}$ & 10$^9$ & 10$^8$ \\
Atomic number & Z& 26 & 26 & 26 \\
Dynamic range & $N^{max}$ & 92 & 7.3$\times$ 10$^4$ &9.2 $\times$10$^5$\\
\hline
\end{tabular}
\end{table}






\section{Single-photon-counting (SPC) modes}\label{sec:SPC}
Both the Cornell-Stanford pixel array detector (CS-PAD) for LCLS-I and the Adaptive Gain Integrating Pixel Detector (AGIPD) for the EXFEL work in the charge integration mode~\cite{CSPAD:2012,AGIPD:2014}; {\it i.e.}, each pixel has a dynamic range around 10$^4$, allowing it to detect one or up to  10$^4$ X-ray photons. These imagers also use high-resistivity silicon as the X-ray converter that turns X-rays into electron-hole pairs as signal.  Here we discuss the motivations for single photon counting (SPC) detectors; {\it i.e.}, each pixel detects no more than one photon for very hard X-rays. We shall call this the strong SPC mode. Then we relax this condition and introduce a weak SPC mode concept. A multilayer 3D detector architecture to realize SPC and weak SPC mode is discussed in the next section.

\subsection{SPC mode: physics and performance considerations}
The motivations for SPC mode of X-ray imaging are a.) The increasingly important Compton scattering fraction in silicon converters for very hard X-rays; and b.) The simultaneous requirements of high detection efficiency (above 50\%) and high speed (1 GHz and higher frame rate).


The integration mode works for CS-PAD and AGIPD because of the relatively small X-ray energy (no more than 12.5 keV). At X-ray energies up to 20 keV, the ratio of Compton scattering to photoelectric absorption is less than 4\% in silicon. However, the ratio increases to 33\% at 40 keV and is no longer negligible. At 100 keV and above (the third harmonic of a 42-keV XFEL is at 126 keV), the ratio becomes 573\% or larger. The integration mode works nicely in low-noise ($<$ 500 e$^-$) silicon sensors since the amount of charge collected is in direct proportion to the number of X-ray photons absorbed. One 12.5 keV X-ray photon will produce 3.4$\times$10$^3$ electrons (3.64 eV per e$^-$/h pair). For very hard X rays as in MaRIE, as well as in some synchrotron facilities such as the Advanced Photon Source at Argonne, the integration mode no longer works as effectively because of the increasing fraction of Compton scattering in silicon.  When an X-ray is Compton scattered, its energy deposition in a pixel is not fixed, but rather distributes over a wide range from zero up to the Compton edge (5.9 keV for 42 keV X-rays). The angular averaged Compton electron energy is 3.1 keV (855 e$^-$). A 42 keV X-ray photon could produce 1.1$\times$10$^4$ e$^-$ (photoelectric absorption) or 1.6$\times$10$^3$ e$^-$ (Compton edge) in silicon. Thus, integrated charge measurement can no longer tell precisely how many photons per pixel through dividing the total charge collected by a constant like 1.1$\times$10$^4$ e$^-$ for 42 keV X-rays. In the SPC mode, measuring the exact amount of charge is not essential. All the signals above a certain threshold, which is limited to the sensor noise, lead to a photon count. Both photoelectric absorption and Compton scattering physics can therefore be accommodated in the SPC mode. Signals below the threshold are lost and result in some loss in efficiency. For a threshold of 500 e$^-$, or the Compton electron energy of 1.8 keV, photons with a Compton scattering angle less than 63$^0$ are not counted. Reducing pixel capacitance can reduce the noise levels and improve the Compton detection efficiency. At a noise level of 50 e$^-$, the Compton scattering angles less than 19$^0$ are not counted. Some other advantages of SPC mode have been discussed previously~\cite{Wang:2012}. 

Detection efficiency requirement determines the sensor thickness. For above 50\% efficiency, at least one X-ray attenuation length would be required. For silicon and 42 keV photons, one X-ray attenuation length is about 7 mm. The sensor thicknesses for silicon and several other semiconductors, which are three times the 42 X-ray attenuation length ($\Lambda_{tot}$) respectively, are shown in Table~\ref{tab:mat}. 

\begin{table}[hbtp]
\caption{Thickness and response time requirements for 42 keV ($\lambda = 0.029$ nm) X-rays.}
\label{tab:mat}
\smallskip
\centering
\begin{tabular}{|llccccc|}
\hline
Material & & C(Diamond) & Si & Ge & GaAs & CdTe \\
Thickness & 3 $\Lambda_{tot}$ (mm) & 42 & 20 & 1.0 & 1.0 &  0.28 \\
Saturated drift speed & v$_d$ ($\mu$m/ns)& 270 & 100& 60 & 200 & $\sim$100 \\
\hline
\end{tabular}
\end{table}

For very thick sensors, when used with the fully depleted diode charge collection configuration, would make GHz frame rate difficult since a.) The charge collection time will increase with sensor thickness; b.) The bias voltage also increases with thickness in order to obtain saturation electric field within the pixel; and c.) potential material issues with very thick sensors. In silicon, for example, electrons move at the saturated drift velocity will take 200 ns to drift a distance of 20 mm. To achieve saturated drift in a thicker sensor will also require a large bias voltage above 10 kV, potentially causing other problems for steady operation. Columnar or 3D deeply entrenched electrode configuration~\cite{Parker:1997}, replacing the existing planar electrode configuration, has recently shown promising results~\cite{Obertino:2013,Cinzia:2013}. It may also be possible that distributed charge collection may be able to separate photoelectrically absorbed and Compton scattered X-rays.

\subsection{The strong SPC mode }\label{sec:sspc}
When $N^{max}$ photons reach a pixel with a thickness $T $, the average number of interactions is given by
\begin{equation}
N_p = N^{max} (\frac{\Lambda_{tot}}{\Lambda_{abs}} + \frac{\Lambda_{tot}}{\Lambda_{inc}})\left[1-\exp (-\frac{T}{\Lambda_{tot}})\right],
\label{eq:spc}
\end{equation}
where $\Lambda_{abs}$ and $\Lambda_{inc}$ are absorption mean free path and the incoherent scattering mean free path respectively. $\Lambda_{tot}^{-1} = \Lambda_{abs}^{-1} + \Lambda_{inc}^{-1} + \Lambda_{coh}^{-1}$. Elastic scattering ($\Lambda_{coh}$) produces no detectable signal. We also ignore the Poisson statistics here. The strong SPC mode is defined as no more than one photon interaction (absorption + inelastic scattering) per pixel, $N_p \leq 1$. The SPC requirement defines the maximum sensor thickness $T_{SPC}$. When $T \leq T_{SPC}$, no more than one X-ray photon on average is detected in each pixel for $N^{max}$. When $N^{max} = 10^3$, $T_{SPC}$ are found to be 7.8, 14, and 0.35 $\mu$m in Si, C(Diamond), Ge/GaAs and 42 keV X-rays, as shown in Fig.~\ref{fig:Layerthick}.

\begin{figure}[htbp] 
\centering
\includegraphics[width=.4\textwidth]{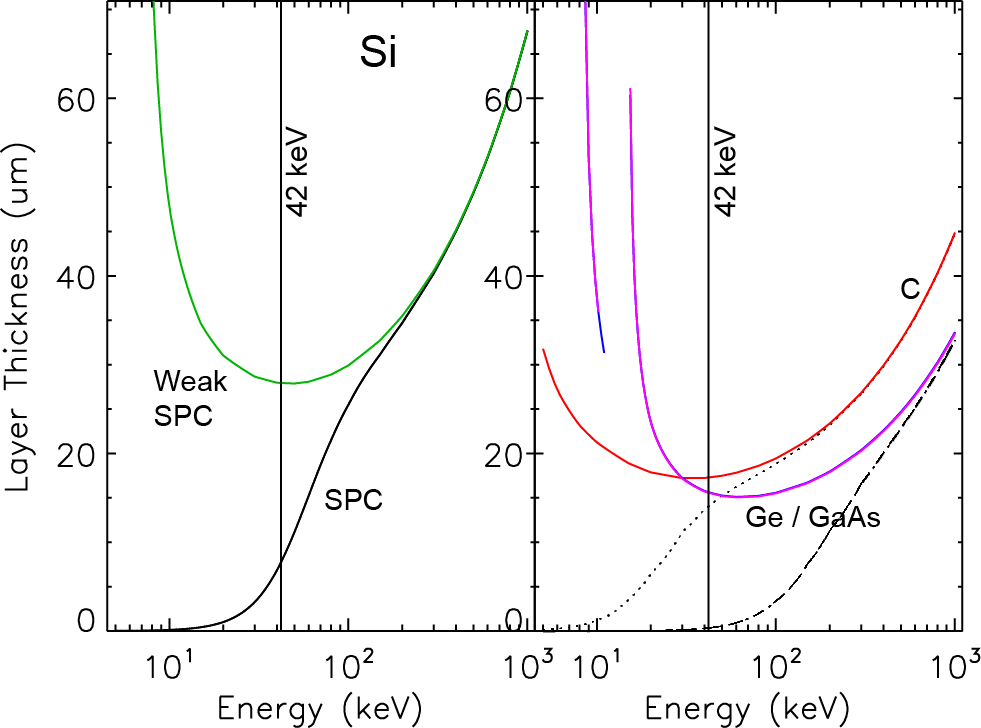}
\caption{Maximum pixel thickness as a function of the X-ray energy, assuming single photon counting (SPC) condition or the weak SPC condition. The X-ray flux onto each pixel is up to $N^{max}=$10$^3$. (Left) the thicknesses for silicon; (Right) The thicknesses for C(diamond), Ge and GaAs (the same).}
\label{fig:Layerthick}
\end{figure}

\subsection{The weak SPC mode}
When $\Lambda_{abs} < \Lambda_{inc}$, there are more absorptions than inelastic scatterings per pixel. we may define a weak SPC mode by allowing multiple absorption and no more than one Compton scattering per pixel. In other words, the weak SPC mode is defined as
\begin{equation}
N_{inc} = N^{max} \frac{\Lambda_{tot}}{\Lambda_{inc}}\left[1-\exp (-\frac{T}{\Lambda_{tot}})\right] \leq 1.
\label{eq:wspc}
\end{equation}
In the weak SPC mode and 42 keV photons, the maximum allowed pixel thickness $T_{SPC}^w$ are found to be 28, 17, and 16 $\mu$m for Si, C, Ge/GaAs respectively. For other energies, the result is shown in Fig.~\ref{fig:Layerthick} along with the result for the SPC mode.

It's worth mentioning that the Compton fraction for 42 keV X-ray photons in Ge and GaAs is only 2.3\%, which may be negligible in many applications. If so, one may be able to use the existing 2D hybrid architecture for high efficiency. Since the total sensor thickness will be still about 1 mm as shown in Table~\ref{tab:mat}, the charge collection time at the saturated electron drift speed will be about 5 ns and 17 ns for Ge and GaAs respectively. If a faster charge collection time of 1 ns is desired, one could stack multiple 2D hybrid sensors on the top of each other. We call this type of multilayer three-dimensional (3D) detector architecture.  For GaAs, five layers of sensors each with a thickness around 200 $\mu$m thick will be needed. For Ge, about 17 sensor layers with a thickness around 60 $\mu$m each. An alternative to the multilayer 3D detector architecture could be a single thick layer with deeply buried electrodes as mentioned above~\cite{Parker:1997,Obertino:2013,Cinzia:2013}. The distances in-between the closest positive and negative electrodes would be less than 200 $\mu$m in GaAs, and less than 60 $\mu$m in Ge. This way, the charge collection time will be around 1 ns or less while maintaining the thick ($\sim$ 1 mm) sensor efficiency. Compared with the multilayer architecture, which will require multiple Application Specific Integrated Circuits (ASICs) to serve multiple layers, one advantage of the 3D electrodes is that one may only need one ASIC to serve the sensor. On the other hand, the aspect ratio of each electrode would be around 1000 for 1 $\mu$m collection electrode radius and 1 mm sensor thickness (corresponds to collection electrode length), which could pose a fabrication challenge.

\section{Multilayer 3D detector architecture}\label{sec:DA}
We further discuss multilayer 3D detector architecture for low-Z materials such as silicon and diamond here. Ge and GaAs based detectors, since their Compton scattering fraction for 40 to 50 keV photons are below 5\%, can use the widely adopted 2D hybrid structures for high efficiency or 3D deeply entrenched electrode structures for high efficiency and large frame rate (above 10 MHz and probably limited by ASIC) simultaneously.

Multilayer 3D detector architecture can be use to achieve SPC or weak SPC mode of detection. Each layer in this architecture is less than 100 $\mu$m so that higher frame-rate, which is limited by the charge collection time, than in existing hybrid detectors such as CS-PAD and AGIPD 1.0 can be achieved. At the saturated electron drift speed, $T_{SPC}$ and $T_{SPC}^w$ correspond to a charge collection time of 78 and 280 ps in Si (100 $\mu$m/ns as listed in Table~\ref{tab:mat}). On the other hand, high efficiency can be obtained by using many layers, making a multilayer detector a 3D detector with hundreds of pixels in all three dimensions. The number of layers or pixels in the third dimension is determined by the ratio of total thickness required for the desired efficiency to individual layer thickness. For a total thickness of 20 mm Si as listed in Table~\ref{tab:mat}, the number of layers are about 2600 and 710 layers for SPC and weak SPC mode.  The total number of pixels would be on the order of 10$^9$, making it a `billion-pixel' detector.

Double-scattering (A Compton scattered photon scatters for the second time in the detector) and therefore duplicated counts in a multilayer 3D detector can be reduced to low levels by adjusting the spacing in-between layers, as shown in Fig.~\ref{fig:Layerthick}. For less than 5\% double counting rate, the layer separation needs to be 2 to 4 mm in C(diamond) and silicon. Since the total number of layers will be order of 1000, the detector size in the third dimension will be about 4 m for a silicon detector, much larger than the lateral dimension of 10 cm (1000 pixel, 100 $\mu$m pixel size). Double scattering is most likely for the forward scattered photons following the first scattering. A more careful examination of double-scattering would be needed, since a balance may be found so that double-scattering events can compensate the loss of efficiency for forward scattered photons, which is not detected during the first scattering because of noise rejection threshold discussed in Sec.~\ref{sec:sspc}.

\begin{figure}[hbtpb] 
\centering
\includegraphics[width=.4\textwidth,angle=90]{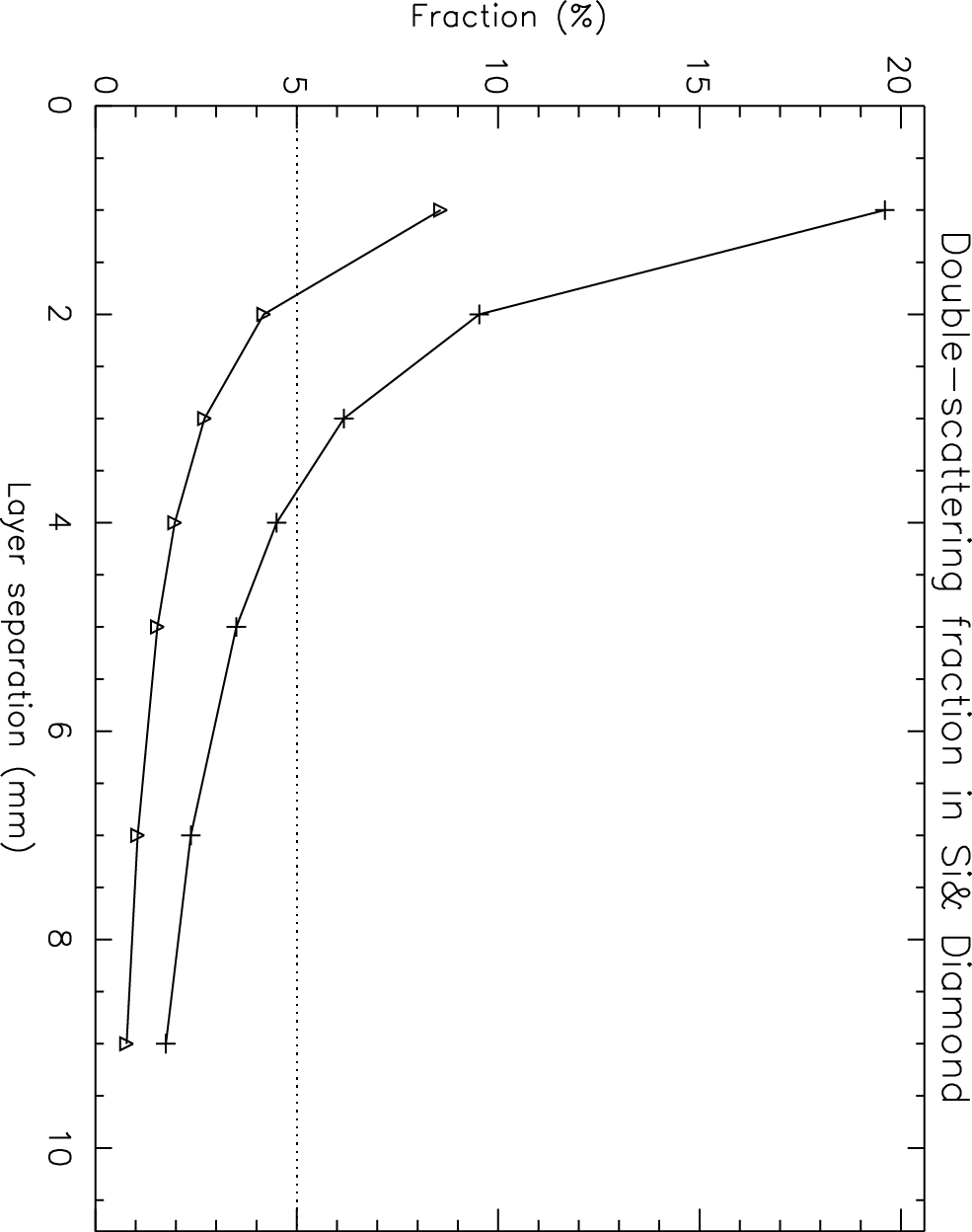}
\caption{The layer-to-layer separation required (abscissa) to reduce double-scattering fractions (ordinate) in a multilayer 3D detector. The minimum spacing in Si (the top curve) is found to be about 4 mm for less than 5\% double-scattering fraction. About 2 mm in C (diamond).}
\label{fig:Layerthick}
\end{figure}

Each layer is a self-sufficient 2D camera by itself, with both sensor and ASIC integrated. the ASIC thickness should be much less than the sensor thickness ($<$ 100 $\mu$m) to avoid efficiency loss in the ASIC, which will be in the path of incoming X-ray photons. Since layer thickness is much less than its lateral dimension and are less than 100 $\mu$m, each layer is essentially a thin-film camera. 

\acknowledgments
Drs. Dinh Nguyen and Richard Sheffield provided inputs about XFEL parameters as summarized in Table~\ref{tab:XFEL}. We also thank Drs. Cris Barnes, Jon Kapustinsky, Chris Morris, Mike Stevens and Jinkyoung Yoo for stimulating discussions and encouragement to carry out the work.


\begin{thebibliography}{9}
\bibitem{Miao:1999} J. Miao, P. Charalambous, J. Kirz and D. Sayre, 
\emph{Extending the methodology of X-ray crystallography to allow imaging of micrometre-sized non-crystalline specimens}, 
{\emph{Nature} {\bf 400} (1999) 342}.

\bibitem{RH:2009}I. Robinson and R. Harder,
\emph{Coherent X-ray diffraction imaging of strain at the nanoscale},
{\emph{Nat. Mater.} {\bf 8} (2009) 291-298}.

\bibitem{Zigler:2011} A. Ziegler,
\emph{Ultrafast material science and 4D imaging with atomic resolution both in space and time},
{\emph{MRS Bull.} {\bf 36} (2011) 121}.

\bibitem{SC:2014} J. C. H. Spence and H. N. Chapman,
\emph{Introduction: The birth of a new field},
{\emph{Phil. Trans. R. Soc. B} {\bf 369} (2014) 2013039}.

\bibitem{Sutton:1991}M. Sutton, S. G. J. Mochrie, T. Greytak, et al.,
\emph{Observation of Speckle by Diffraction with coherent X-rays},
{\emph{Nature} {\bf 352}(6336) (1991) 608-610}.

\bibitem{Wilkins:1996}S. W. Wilkins, T. E. Gureyev, D. Gao, A. Pogany and A. W. Stevenson,
\emph{Phase-contrast imaging using polychromatic hard X-rays},
{\emph{Nature} {\bf 384} (1996) 335}.

\bibitem{Luo:2012} S. N. Luo, B. J. Jensen, D. E. Hooks, et al.,
\emph{Gas gun shock experiments with single-pulse X-ray phase contrast imaging and diffraction at the Advanced Photon Source},
{\emph{Rev. Sci. Instrum.} {\bf 83} (2012) 073903}. 

\bibitem{HD:1955} B. L. Henke and J. W. M. DuMond,
{\emph{J. Appl. Phys.} {\bf 26} (2003) 2344}.

\bibitem{Marchesini:2003} S. Marchesini, H. N. Chapman, S. P. Hau-Riege et al., 
{\emph{Opt. Express} {\bf 11} (2003) 903-917}.

\bibitem{Shen:2004} Q. Shen, I. Bazarov and P. Tribault,
\emph{Diffractive imaging of nonperioidic materials with future coherent X-ray sources},
{\emph{J. Synchrotron Rad.} {\bf 11} (2004) 432}.

\bibitem{Howells:2009} M. R. Howells, et al.,
\emph{An assessement of the resolution limitation due to radiation-damage in X-ray diffraction microscopy},
{\emph{J. Electron Spectr. Rel. Phen.} {\bf 170} (2009) 4}.

\bibitem{Barber:2014}J. L. Barber, C. W. Barnes, R. L. Sandberg and R. L. Sheffield,
\emph{Diffractive imaging at large Fresnel number: Challenge of dynamic mesoscale imaging with hard X-rays},
{\emph{Phys. Rev. B.} {\bf 89} (2014) 184105}. 

\bibitem{Gruner:2002} S. M. Gruner, M. W. Tate, and E. F. Eikenberry,
{\emph{Rev. Sci. Instrum.} {\bf 73} (2002) 2815}.

\bibitem{Falus:2004} P. Falus, M. A. Borthwick and S. G. J. Mochrie,
{\emph{Rev. Sci. Instrum.} {\bf 75} (2004) 4383}.

\bibitem{CSPAD:2012} P. Hart, et al.,
\emph{Proc. SPIE} {\bf 8504} (2012) 85040.

\bibitem{AGIPD:2014}
A. Allahgholi et al., \emph{AGIPD, a high dynamic range fast detector for the European XFEL}, \jinst{10}{2014}{C01023}.

\bibitem{nxd:rpt}G. Carini, P. Denes, S. Gruner, and E. Lessner,
\emph{Neutron and X-ray detectors},
\url{http://science.energy.gov/~/media/bes/pdf/reports/files/NXD_rpt.pdf}
Aug. 1-3, (2012).

\bibitem{pri:Dinh} Dinh C. Nguyen and Richard L. Sheffield (private communications).

\bibitem{Arthur:2002} J. Arthur, et al., 
\emph{Linac Coherent Light Source (LCLS) Conceptual Design Report},
SLAC-R-593, UC-414 (2002).

\bibitem{Altarelli:2007} M. Altarelli, et al., 
\emph{The European X-Ray Free-Electron Laser, Technical design report}
DESY 2006-097 (2007).

\bibitem{PS}
C. Pellegrini and J. St\"ohr,
\emph{X-ray Free Electron Lasers: Principles, Properties and Applications},
\url{http://ssrl.slac.stanford.edu/stohr/xfels.pdf}

\bibitem{Warren:1990}
B. E. Warren, \emph{X-ray diffraction}, Dover (1990).

\bibitem{Guinier:1994}
A. Guinier, \emph{X-ray diffraction in crystals, imperfect crystals and amorphous bodies}, Dover (1994).

\bibitem{Sayre:1980}
D. Sayre, in \emph{Imaging Processes and Coherence in Physics}, M. Schlenker (ed.s), Springer-Verlag, Berlin (1980).

\bibitem{Miao:1998}
J. Miao, D. Sayre and H. N. Chapman, \emph{J. Opt. Soc. Am.} A {\bf 15}, 1662-1669 (1998).

\bibitem{NIST} \url{http://physics.nist.gov/PhysRefData/Star/Text/ESTAR.html}

\bibitem{Wang:2012} Z. Wang, C. L. Morris, J. S. Kapustinsky, K. Kwiatkowski and S.-N. Luo,
{\emph{Rev. Sci. Instrum.} {\bf 83} (2012) 10E510}.

\bibitem{Parker:1997}S. I. Parker, C. J. Kenney and I. Segal,
{\emph{Nucl. Instrum. Meth. A.} {\bf 395}, 328 (1997)}.

\bibitem{Obertino:2013} M. Obertino, et al.,
{\emph{Nucl. Instrum. Meth. A.} {\bf 718}, 342 (2013)}.

\bibitem{Cinzia:2013}C. Da Vi\`a, et. al., 
{\emph{Nucl. Instrum. Meth. A.} {\bf 731}, 201 (2013)}.




\end{thebibliography}
\end{document}